\documentclass[prc,showpacs]{revtex4}

\usepackage{graphicx}
\usepackage{dcolumn}
\usepackage{bm}
\newcommand {\car}{$^{12}$C }
\newcommand {\oxi}{$^{16}$O }
\newcommand {\caI}{$^{40}$Ca }
\newcommand {\caII}{$^{48}$Ca }
\newcommand {\lead}{$^{208}$Pb }
\newcommand{\br}{{\bf r}}
\newcommand{\half} {\frac{1}{2}}
\newcommand{\bsigma}{\mbox{\boldmath $\sigma$}}
\newcommand{\btau}{\mbox{\boldmath $\tau$}}
\begin{document}

\title{Ground state of medium-heavy doubly-closed shell nuclei \\
in correlated basis function theory}
\author{C. Bisconti$^{\,1}$,
F.Arias de Saavedra$^{\,2}$, G.Co'$^{\,1}$ and A. Fabrocini$^{\,3}$}
\affiliation{
$^1$
Dipartimento di Fisica, Universit\`a di Lecce \\
 and Istituto Nazionale di
Fisica Nucleare, sezione di Lecce,  \\  
I-73100 Lecce, Italy \protect\\
$^2$ Departamento de F\'{\i}sica Moderna, 
Universidad de Granada, \\
E-18071 Granada, Spain \protect\\
$^3$ Dipartimento di Fisica, Universit\`a di Pisa,\\ 
and Istituto Nazionale di Fisica Nucleare, sezione di Pisa,\\
I-56100 Pisa, Italy 
}

\date{\today}

\begin{abstract}
The correlated basis function theory is applied to the study of
medium-heavy doubly closed shell nuclei with different wave functions
for protons and neutrons and in the $jj$ coupling scheme. State
dependent correlations including tensor correlations are
used. Realistic two-body interactions of  Argonne and Urbana type,
together with three-body interactions have been used to calculate
ground state energies and density distributions of the \car, \oxi,
\caI, \caII and \lead nuclei.
\end{abstract}

\maketitle

\section{INTRODUCTION}
\label{sec:int}
In the last decade, the validity of the non relativistic many-body
theory in the description of nuclear systems, has been well
established. The idea is to describe the nucleus as a set of mutually
interacting nucleons with a hamiltonian of the type
\begin{equation}
H = \sum_i \frac{-\hbar^2}{2m_i}\nabla^2_i + \sum_{i<j}v_{ij}
 +\sum_{i<j<k}v_{ijk} 
\label{eq:hami}
\end{equation}
where the two- and three-body interactions, $v_{ij}$ and $v_{ijk}$
respectively, are fixed to reproduce the properties of the two- and
three-body systems.

Quantum Monte Carlo calculations solve the many-body Schr\"odinger
equation without approximations \cite{pud95,pud97} and have been
applied with success to describe nuclei up to twelve nucleons
\cite{pie05}. The only limitation of this approach is related to the
available computing power, which, at present, does not allow its
straightforward application to medium and heavy nuclei.  The good
results obtained by the Quantum Monte Carlo calculations have renewed
the interest in nuclear structure calculations with realistic, or
microscopic, interactions.  For this reason, many-body techniques,
which are not approximation free, have been formulated, or
reconsidered, and applied to the study of medium and heavy nuclei, or
more in general, to nuclear systems with A$>$4.

The oldest, and most commonly used, technique is the Brueckner
Hartree-Fock approach recently applied to calculate the properties of
\oxi \cite{sch91,dic04}. In the last years, no-core shell model
calculations \cite{nav00,for05} have used microscopic interactions to
calculate some property of nuclei up to \car. The coupled cluster
method has been used to evaluate \oxi properties
\cite{hei99,mih99,dea05}.  More recently a technique called Unitary
Correlation Operator Method has been developed to produce effective
interactions used in mean-field and shell model calculations, and it
has been applied with success to a variety of nuclei up to \lead
\cite{nef03,rot04}.

In the early nineties, we started a project aimed to apply to finite
nuclei the Correlated Basis Function (CBF) theory, originally
developed for infinite systems \cite{ros82}.  The idea was to use the
Fermi Hypernetted Chain (FHNC) resummation technique, well tested in
nuclear matter, to describe finite systems.  We first tested the
validity of our approach in model nuclei with degenerate proton and
neutron states, and in $ls$ coupling scheme \cite{co92,co94}.  Then,
by using central interactions and scalar correlation functions, we
applied the theory to describe nuclei with different single particle
bases for protons and neutrons, with different number of protons and
neutrons and with the more realistic $jj$ coupling scheme
\cite{ari96}.  More recently, we have treated two-body interactions
containing tensor and spin-orbit terms and this implies the use of
state dependent correlations \cite{fab98}. This was not a
straightforward extension of the formalism because the non
commutativity of the different operators requires the use of the
Single Operator Chain (SOC) approximation in the FHNC resummation.
The FHNC/SOC computational scheme has been extended to include
spin-orbit and three-body interactions of Urbana and Argonne type
\cite{fab00}.  In these last calculations protons and neutrons single
particle wave functions were distinguished only by the third component
of the isospin, and they were treated in the $ls$ coupling scheme.
This choice allowed us to treat only systems saturated in spin and
isospin.

This paper represents a step forward in our project.  The
FHNC/SOC formalism has been extended to treat nuclei with different
proton and neutrons wave functions, and in a $jj$ coupling scheme in
spherical basis.  With this single particle basis we obtained results
for the \car, \oxi, \caI, \caII and \lead nuclei.

The aim of this paper is to present the results we have obtained by
using the two-body Argonne $v'_8$ \cite{wir95,pud97} and the Urbana
$v_{14}$ \cite{lag81a,lag81b} interactions, where we have considered
all the channels up to the spin-orbit ones. In our calculations we
have implemented these two-body interactions with the Urbana IX
\cite{pud97} and the Urbana VII \cite{sch86} three-body interactions
respectively. 

In Sect. \ref{sec:for} we briefly discuss the basic ideas of the
formalism pointing out the main differences with respect to what has
been presented in \cite{fab98} and \cite{fab00}.  The detailed
presentation of the heavy formalism, which is given in \cite{bis05},
is beyond the scope of this article.

\section{THE FORMALISM}
\label{sec:for}
Our approach is based on the variational principle, i.e. we search
for the minimum of the energy functional
\begin{equation}
E[\Psi]= \frac{<\Psi|H|\Psi>}{<\Psi|\Psi>}
\label{eq:efun}
\end{equation}
We used 
the hamiltonian of eq. (\ref{eq:hami}) in our calculations,
and we express the two-body interaction in the form
\begin{equation}
v_{ij}=\sum_{p=1,8} v_p(r_{ij}) O^p_{ij} 
\label{eq:vij}
\end{equation}
where we have defined the operators $ O^p$ as:
\begin{equation}
 O^{p=1,8}_{ij} = [1,\bsigma_i \cdot \bsigma_j,S_{ij},
({\bf L}\cdot{\bf S})_{ij}]\otimes[1,\btau_i\cdot\btau_j] 
\label{eq:operator}
\end{equation}
and where we have indicated with $S_{ij}$ the tensor operator
\begin{equation}
S_{ij}= 3 \frac{\br_{ij}\cdot \bsigma_i \br_{ij}\cdot \bsigma_j}
               {r^2_{ij}} - \bsigma_i \cdot \bsigma_j
\end{equation}
In eq. (\ref{eq:vij}) we have limited the sum to the first 8 channels
which are those used in our calculations. In reality modern two-body
interactions have been parametrized by using up to 18 different
channels \cite{wir95}.

The three-body interaction \cite{pud97,sch86} is composed by two
parts. A first one describes the process where two pions are exchanged
with the intermediate excitation of a $\Delta$ \cite{fuj57}. This
term describes the long-range part of the three-body interaction and,
usually, gives an attractive contribution. The second part of the
three-body interaction is a shorter range, repulsive, spin and isospin
independent term. This term is supposed to simulate dispersive effects
obtained when the $\Delta$ degrees of freedom are integrated out.

The search for the energy minimum (\ref{eq:efun}) is done in the
subspace of the many-body wave functions $\Psi$ which can be expressed 
as:
\begin{equation}
\Psi(1,2,...,A)=F(1,2,...,A)\Phi(1,2,....,A)
\end{equation}
where $F(1,2,...,A)$ is a many-body correlation function and 
$\Phi(1,2,....,A)$ is a Slater determinant composed by a set of single
particle wave functions $\phi_\alpha(i)$.
In our calculations we describe the many-body correlation function as
a symmetrized product of two-body correlation functions:
\begin{equation}
F(1,2,...,A)={\mathcal S} \left[ \prod_{i<j} F_{ij} \right]
\end{equation}
and we use two-body correlation functions $F_{ij}$ which have an
operator dependence analogous to that of the nucleon-nucleon
interaction (\ref{eq:vij}), but without the spin-orbit terms:
\begin{equation}
F_{ij} = \sum_{p=1,6} f_p(r_{ij}) O^p_{ij} 
\end{equation}
\subsection{The cluster expansion}
\label{ssec:cluster}
The evaluation of the multidimensional integrals necessary to
calculate the energy functional 
(\ref{eq:efun}) is done by performing a cluster
expansion and by applying the FHNC resummation techniques.  Our work
is based on the formalism developed in \cite{pan79,co92,ari96,fab98}.

The two basic quantities  used in the cluster expansion of
eq. (\ref{eq:efun}) are the one- and the 
two-body density matrices respectively defined as:
\begin{eqnarray}
\rho_1(x_1,x'_1) &=& A \int dx_2\,dx_3...dx_A \,\Psi^*(x_1,x_2,...,x_A)
                                     \Psi(x'_1,x_2,...,x_A)
  /<\Psi|\Psi>
\label{eq:obdc}\\
\rho_2(x_1,x_2) &=& A(A-1)
\int dx_3...dx_A \,\Psi^*(x_1,x_2,x_3,...,x_A)
                                     \Psi(x_1,x_2,x_3,...,x_A)
 /<\Psi|\Psi>
\label{eq:tbdc}
\end{eqnarray}
where we have indicate with $x_i$ the coordinate fully characterizing
the particle $i$, i.e. position $\br_i$, 3-rd components of the spin,
$s$, and of the isospin, $t$. The evaluation of the expectation values
of one- and two-body operators can be obtained by making the operators
acting on the densities (\ref{eq:obdc}) and (\ref{eq:tbdc}) and
integrating on the free variables.

In the uncorrelated case, $F(1,2,...,A)=1$, the expressions
(\ref{eq:obdc}) and (\ref{eq:tbdc}) become:
\begin{eqnarray}
\rho^o_1(x_1,x'_1) &=& \sum_i \phi^*_i(x_1) \phi_i(x'_1)
\label{eq:obdu}\\
\rho^o_2(x_1,x_2) &=& \rho^o_1(x_1)\rho^o_1(x_2) -  
                    \rho^o_1(x_1,x_2)\rho^o_1(x_2,x_1)
\label{eq:tbdu}
\end{eqnarray}
where we have indicated with $\phi_i$ the single particle wave
functions.  In the above equations, the sum is done on all the
occupied states, and we have used the notation $\rho^o_1(x_1) \equiv
\rho^o_1(x_1,x_1)$ to indicate the diagonal part of the one-body
density function.

We construct the single particle wave functions by using a spherical
mean field potential containing a spin-orbit term. Protons and
neutrons wave functions are generated by different potentials. The
single particle wave functions are classified in terms of the angular
momentum $j$ and of the third component of the isospin $t$. They can
be written as:
\begin{equation}
\phi^t_{nljm}(\br_i) = R^t_{nlj}(r_i) \sum_{\mu,s} <l \mu \half s| j m> 
Y_{l\mu}(\Omega_i) \chi_s(i)\chi_t(i)
\label{eq:spwf}
\end{equation}
where we have indicated with $Y_{l\mu}(\Omega_i)$ the spherical harmonics, 
with $\chi_{s,t}(i)$ the Pauli spinor/isospinor and with 
$<l \mu \half s| j m>$ the Clebsch-Gordan coefficient.

In the calculations of Refs. \cite{co92,fab98,fab00}, the radial part
of the single particle wave function was independent on the isospin.
This, joined to the fact that we studied doubly closed shell nuclei
on $ls$ coupling with the same number of protons and neutrons, 
allowed us to separate
the spatial part from the spin and isospin dependent parts in
$\rho^{o}_1(x_1,x'_1)$. Because we restricted our investigation to 
finite systems saturated in spin and isospin, we could use the trace
techniques developed in \cite{pan79} to deal with state dependent
correlations.

In the present case, as pointed out in \cite{ari96}, the choice of
the single particle basis (\ref{eq:spwf})
requires to consider the spin dependence of
the radial part of the one-body density matrix:
\begin{equation}
\rho^{o}_1(x_1,x'_1) = \sum_{s,s',t} \rho^{ss' t}_0(\br_1,\br'_1)
\chi^\dagger_s(1) \chi_{s'}(1')
\chi^\dagger_t(1) \chi_{t}(1')
\label{eq:rhossp}
\end{equation}
therefore the factorization between spatial and spin-isospin
dependence of the one-body density matrix is lost.

We consider two radial terms of the one-body density matrix. A first one,
$\rho_0^t$, related to the parallel spin terms, and the other one,
$\rho_{0j}^t$, to the antiparallel spin terms. We can rewrite eq.
(\ref{eq:rhossp}) as:
\begin{equation}
\rho^{o}_1(x_1,x'_1) = \hspace*{-1mm}
\sum_{t} \chi^\dagger_t(1) \chi_{t}(1')
\Bigg( 
\rho_0^t(\br_1,\br'_1) \sum_{s} \chi^\dagger_s(1) \chi_{s}(1') +
\rho_{0j}^t(\br_1,\br'_1) \sum_{s} (-1)^{s-1/2}\chi^\dagger_s(1) \chi_{-s}(1')
\Bigg)
\label{eq:rhossp1}
\end{equation}
The explicit expressions of $\rho_0^t$ and $\rho_{0j}^t$, in terms of
single particle wave functions, are given in \cite{ari96}. 

The study of Ref. \cite{ari96} has shown that the main contribution to
the one-body density matrix is given by $\rho_0^t$, which is the only
term surviving in spin saturated systems having degenerate spin-orbit
partners. The similarity of this term to the one appearing in the $ls$
coupling scheme, allows us to use some simplifications in the
calculations of the correlated distributions
(\ref{eq:obdc},\ref{eq:tbdc}) following the procedure adopted in
\cite{pan79}.  Even though the contribution of the $\rho_{0j}^t$ terms
are much smaller than those of $\rho_{0}^t$, they cannot be neglected,
as it is indicated by the sum rule results. For this reason, we had to
calculate new kinds of spin traces.

In the calculation of the one-body density matrix and of the two-body
density function (\ref{eq:obdc}) and (\ref{eq:tbdc}) we consider the
scalar terms of the correlations at all the orders of the cluster
expansion. This part of the calculation follows step by step that
presented in \cite{ari96}.  The two-body densities are written in
terms of the scalar correlation $f_1(r)$, of the {\sl nodal}, or {\sl
  chain}, diagrams $N_{xy}(\br_1,\br_2)$, of the non nodal (composite) 
diagrams $X_{xy}(\br_1,\br_2)$ and of the {\sl elementary}, or {\sl
  bridge}, diagrams $E_{xy}(\br_1,\br_2)$. Here the two sub-indexes $x$
and $y$ identify the exchange character of the external points $1$ and
$2$, as classified in \cite{co92}.

The situation becomes more complicated when state dependent
correlations are introduced, essentially because the various terms of
the correlations do not commute with the hamiltonian and among
themselves.  At present, a complete FHNC treatment for the full, state
dependent cluster expansion of the two-body densities is not
available. In our calculations we adopted the Single Operator Chain
(SOC) approximation \cite{pan79} consisting in summing those operator
chains containing scalar dressing at all orders, but only a single
operator $p>1$ element.

The different kind of diagrams: nodal, non-nodal  
and elementary ones, are characterized
by an index $p$ denoting the operator (\ref{eq:operator})
associated to the correlation 
function. In order to treat nuclei with different
number of protons and neutrons ($N \ne Z$) we separate the isospin 
dependence of the correlation operators from that of the other operators:
\begin{equation}
F(i,j)=\sum_{k=1}^{3}\sum_{l=0}^{1}
f_{2k-1+l}(r_{ij})O^{2k-1+l}_{ij}=
\sum_{l=0}^{1}(\btau_{i}\cdot\btau_{j})^{l}
\sum_{k=1}^{3}f_{2k-1+l}(r_{ij})P_{ij}^{k}
\end{equation}
with $P_{ij}^{k=1,2,3}=1,\bsigma_{i}\cdot\bsigma_{j},S_{ij}$.

With respect to the treatment presented in \cite{fab98}, now we have
to consider that the various diagrams depend on the third isospin
component of the nucleons located in the two external points. As a
consequence, the number of cluster diagrams to be calculated is four
that of the previous cases. In reality, for symmetry reasons, we had
to calculate separately only three different types of diagrams.  In
addition, we should take into account also the contribution of the
uncorrelated two-body densities with antiparallel spins, the terms
related to $\rho_{0j}^t$ in eq.  (\ref{eq:rhossp1}).  The
contributions of these two-body densities are not zero only for the
scalar, $p=1$, and isospin, $p=2$, operators. As we have already
mentioned in the introduction, a detailed presentation of all the
expressions used in our calculations can be found in \cite{bis05}.

\subsection{The energy expectation value}
The calculation of the energy expectation value has been done along
the lines of Refs. \cite{ari96} and \cite{fab98}.  We used the
Jackson-Feenberg identity to calculate the kinetic energy, since this
expression allows us to eliminate terms of the type $(\nabla_i
F)(\nabla_i \Phi)$ involving three-body operators \cite{fan79}.
Following the notation of Ref. \cite{co92}, we express the kinetic
energy by separating those terms where the $\nabla$ operator acts on
the correlation function $F$ only, from those where it acts on the
uncorrelated many-body wave function $\Phi$.
\begin{equation}
\frac {\langle \Psi | T | \Psi\rangle} {\langle \Psi| \Psi \rangle}
= T_{JF} = T_\phi + T_F ,
\label{eq:tjf}
\end{equation}
where we have defined \cite{co92}:
\begin{equation}
 T_\phi=-A{\frac{\hbar^2}{4m}}\langle \Phi^* F^2 \nabla_1^2 \Phi 
-  \left( \nabla_1 \Phi^*\right) F^2 \left( \nabla_1 \Phi \right) \rangle 
\label{eq:tphi}
\end{equation}
and 
\begin{equation}
T_F=-A{\frac{\hbar^2}{4m}}\langle \Phi^* 
\left[  F \nabla_1^2 F - \nabla_1 F \cdot  \nabla_1 F \right]
 \Phi \rangle , 
\label{eq:tf}
\end{equation}

The operator structure of the $T_F$ terms is similar to that obtained
by inserting the two-body interaction term of the hamiltonian. For
this reason, we found convenient to calculate the joint contribution of
the $T_F+V_2=W$ operator, called {\sl interaction energy} in
\cite{fab98}. In this calculation we have considered all the
interaction terms up to $p=6$. 
The expectation value of $W$ can be expressed in terms of
\begin{eqnarray}
\nonumber
&~&
H_{JF}^{p,q,r}(r_{12}) \equiv
W^{p,q,r}(r_{12}) = T^{p,q,r}_F(r_{12})+V_2^{p,q,r}(r_{12}) \\
\nonumber
&~&
\hspace*{-2mm} =
\frac{1}{(f^{1}(r_{12}))^2}\biggl(-\frac{\hbar^2}{2m}\delta_{q1}
\left\{ f^p(r_{12})\nabla^2 f^r(r_{12})-
 \nabla f^p(r_{12})\cdot\nabla f^r(r_{12})\right\}
 +f^p(r_{12})v^q(r_{12})f^r(r_{12})\biggr) \\
&~&
\label{eq:HJF}
\end{eqnarray}
where the $p,q,r$ labels refer to the type of operator and can assume
values from 1 to 6. Following what has been done in \cite{fab98},
we found convenient to split $W$ in four parts identified by the
different topology of the various links,
\begin{equation}
W=W_0+W_s+W_c+W_{cs},
\label{eq:w}
\end{equation}
The $W_0$ term is the sum of the diagrams having only scalar chains
between the interacting points, connected by $H_{JF}$. We indicate
with $W_s$ the sum of the diagrams with a Single Operator Ring (SOR)
touching the interacting points and scalar chains. The $W_c$ term
contains diagrams with a Single Operator Chain (SOC) between the
interacting points and $W_{cs}$ contains both SOR and SOC between the
interacting points.  The expressions we obtain for the contributions
of the various $W$ terms are more involved than those presented in
\cite{fab98}. This because we have to sum on all the isospin indexes
and, in addition, we must consider also the contribution of the exchange
links with different spin projections. The detailed expressions are
given in \cite{bis05}. In reality, we calculate explicitly the $W_0$,
$W_s$ and $W_c$ contributions, while for the $W_{cs}$ we used the
approximation $W_{cs} \sim W_c[W_s/W_o]$ whose validity is discussed
in \cite{fab98}.

As an example of the differences between the present calculations and
those of \cite{fab00}, we compare the expression of $W_0$ in $ls$
coupling scheme and $N=Z$ that is
\begin{equation}
W_0 = \frac 1 2 \int d{\bf r}_1 d{\bf r}_2
H_{JF}^{p,q,r}(r_{12}) \biggl[ 
K^{pqr} A^{r} \rho_{2,dir}(1,2)+
\Delta^{n} K^{npm} K^{qrm} A^m \rho_{2,exc} (1,2) \biggr] 
\label{eq:w01}
\end{equation}
with the present expression of $W_0$ given by
\begin{eqnarray}
W_0 & = & \frac 1 2 \int d{\bf r}_1 d{\bf r}_2
H_{JF}^{2k_1-1+l_1,2k_2-1+l_2,2k_3-1+l_3}(r_{12}) \biggl[ 
I^{k_1 k_2 k_3} A^{k_3} 
\rho_{2,dir}^{\alpha \beta} (1,2) 
\chi_{l_1+l_2+l_3}^{\alpha \beta} +\nonumber \\
& & \hspace*{-1.2cm} \Bigl( 
\chi_{l_1+l_2+l_3}^{\alpha \beta} +
\chi_{l_1+l_2+l_3+1}^{\alpha \beta} \Bigr)
\Delta^{k_4} I^{k_4 k_1 k_5} I^{k_2 k_3 k_6} \Bigl( 
I^{k_5 k_6 1} \rho_{2,exc}^{\alpha \beta} (1,2) +
I^{k_5 k_6 2} \rho_{2,excj}^{\alpha \beta} (1,2) \Bigr) 
\biggr] . 
\label{eq:w02} 
\end{eqnarray}
In both equations, a sum on repeated indexes is understood, with
$p,q,r,m,n=1,\ldots,6$ in (\ref{eq:w01}) and $\alpha,\beta=p,n$, $k_i
=1,2,3$ and $l_i=0,1$ in (\ref{eq:w02}). We have indicated with
$\rho_{2,dir}$ the direct part of the two-body density function
which includes all the diagrams where the two external particles, 1
and 2, do not belong to the same statistical loop. 
Eq. (\ref{eq:w01}) has only one  exchange part
of the two--body density function , $\rho_{2,exc}$. On the
contrary eq. (\ref{eq:w02}) contains two terms depending weather the 
spin of the external particles are parallel, 
$\rho_{2,exc}^{\alpha \beta}$, or antiparallel,
$\rho_{2,excj}^{\alpha \beta}$.

While the two-body density functions are isospin independent in eq.
(\ref{eq:w01}), in the present calculations we have to
distinguish the cases when the external particles are protons or
neutrons. This isospin dependence obliges us to isolate the isospin
traces from those related to the other operators. These traces
can be written in terms of
\begin{equation}
\chi_n^{\alpha \beta} = \chi_{\alpha}^* (1)\chi_{\beta}^* (2)
\left(\btau_1 \cdot \btau_2 \right)^n \chi_{\alpha} (1)
\chi_{\beta} (2) 
\end{equation}
In eq. (\ref{eq:w01}) the isospin trace is included in the $K$
matrices, defined in \cite{pan79}.  The $I$ matrices of eq.
\ref{eq:w02} are the spin traces of the products of operators,
therefore, they are included in $K$.  As we have already mentioned,
the presence of a new statistical link in the exchange part of eq.
(\ref{eq:w02}) requires the evaluation of spin traces absent in the
$ls$ coupling case.

The differences pointed out in the previous discussion are present in
all the terms of the kinetic energy. To simplify the calculations, we
have calculated the contribution of $\rho_{2,excj}^{\alpha \beta}$
only in the $W_0$ and $W_s$ parts of the interaction energy. In these 
two terms the contribution of $\rho_{2,excj}^{\alpha \beta}$ is about
two orders of magnitude smaller than that of $\rho_{2,exc}^{\alpha
  \beta}$.  We have estimated that this relation between the two
exchange terms is not going to be modified in $W_{c}$.

The calculation of the kinetic energy contribution requires the
evaluation of the $T_\phi$ term of eq. (\ref{eq:tphi}). Also in this
case, in analogy to what we have done in \cite{ari96,fab98}, we
separate the term in three parts:
\begin{equation}
T_\phi= T_\phi^{(1)} + T_\phi^{(2)} +  T_\phi^{(3)} .
\label{eq:tphi123}
\end{equation}
where the term $n=1$ corresponds to the contributions where the
kinetic energy operator acts on a nucleon not involved in any exchange,
the term $n=2$ when it belongs to a two-body exchange loop, and  $n=3$
to a many-body exchange loop.

The expression obtained for $T_\phi^{(1)}$ is the same as that given
in \cite{fab98}. The other terms are more involved because of their
dependence on the isospin indexes, the presence of new isospin traces
and of the exchange links with the $\rho^o_j$ uncorrelated
densities \cite{bis05}.

The evaluation of the {\sl interaction energy} (\ref{eq:HJF}), allows
us to calculate all the two-body interaction channels up to $p=6$. The
contribution of the remaining terms of the two-body interaction, the
$p=7,8$ spin-orbit channels, has been calculated following the
procedure and the approximations discussed in \cite{fab00}. We 
calculate only the $W_0$ term with the $f_6$
correlation function.  Again, the expressions we have obtained are
more involved than those given in \cite{fab00} since we have isospin
dependent terms, and we have considered the contribution of the
leading $\rho^o_j$ in the exchange terms.

The above discussion can be extended to describe our treatment of the
three-body interaction. We have used the nuclear matter results of
Ref. \cite{car83} to select the relevant diagrams to be evaluated.  We
followed the lines of calculations presented in \cite{fab00} and we
took care of the new aspects caused by the $jj$ coupling and $N\ne Z$ 
\cite{bis05}.

\section{RESULTS}
\label{sec:res}
The FHNC/SOC formalism presented above, has been applied to calculate
binding energies and matter distributions of \car, \oxi, \caI, \caII,
and \lead doubly magic nuclei. A complete variational calculation
would imply the search for the energy minimum by modifying both the
set of single particle wave functions forming $\Phi$, and the
correlations function $F$. These calculations are numerically very
involved since the number of the free parameters to be modified is
rather large, about ten.  To reduce the computational effort in our
calculations we kept fixed the single particle wave functions and we
made variations on the correlation function only.  At first sight this
can appear a severe limitation, however we have shown in \cite{fab00}
that large changes of the mean field wave functions produce small
variations of the value of the total energy of the system. For this
reason, by extrapolating the findings of \cite{fab00} we are confident
that our results are only few percent above the minimum obtained with
a full minimization.

In our calculations we adopted the same set of single particle wave
functions used in \cite{ari96}. They are generated by a Woods-Saxon
central potential containing a spin-orbit term. Different potentials
have been used for protons and neutrons. The parameters of these
potentials, given in \cite{ari96}, have been taken from the literature
where they have been fixed to reproduce, at best, the single particle
energies close to the Fermi energy, and the charge density
distributions.

The other fixed input of our calculations are the nucleon-nucleon
interactions.  In this article we discuss mainly the results obtained
with the two-body Argonne $v'_{8}$ interaction \cite{wir95,pud97}
implemented with the Urbana IX ($UIX$)
three-body force \cite{pud97}.  As a
comparison, we discuss also results obtained with the Urbana $v_{14}$
two-body interaction \cite{lag81a,lag81b} but considering only the
first eight channels. Together with the $v_{14}$ two-body interaction
interaction we used the Urbana VII ($UVII$) \cite{sch86}
three-body interaction.

The hamiltonian and the single particle basis are fixed input of our
calculations and the variations have been done on the correlation
function $F$.  The two-body correlation functions $f_p(r)$ are the
functions we have modified in order to find the minimum of the energy
functional. We fixed these functions by using the technique we have
defined as {\sl Euler procedure} in our previous works
\cite{ari96,fab98}. We solve the FHNC/SOC equations up to the second
order in the cluster expansion by forcing the asymptotic behaviour of
the two-body correlation functions $f_p(r)$. More specifically we
imposed that, after certain asymptotic distances, that we call {\sl
  healing distances}, the scalar two-body correlations ($p=1$) must be
equal to one, while the other correlation functions must be zero. The
values of the {\sl healing distances} are our variational parameters.

In the actual calculations, in order to reduce the number of
variational parameters, we used only two {\sl healing distances}. A
first one, $d_c$, for all the central channels, $p \le 4$, and a second
one, $d_t$, for the two tensor channels $p =5,6$.  Nuclear matter
\cite{pan79} and finite nuclei \cite{fab98,fab00} calculations
indicate that tensor and central correlations {\sl heals} at
different distances. Minima are found for tensor {\sl healing
distances} larger than the central ones.

In Tab. \ref{tab:heal}, we present the values of the healing distances
giving the energy minima for each nucleus, and for both interactions
$v'_8$ with $UIX$ and $v_{14}$ with $UVII$. It is remarkable the
similarity of these values for all the nuclei but the \car nucleus.
The values of $d_c$ for the two interactions are the same, with the
usual exception of \car, while the $d_t$ values for $v_{14}$ plus
$UVII$ are slightly larger than those obtained with $v'_8$ plus $UIX$.
 
\begin{figure}[ht]
\includegraphics [scale=0.45]
                 {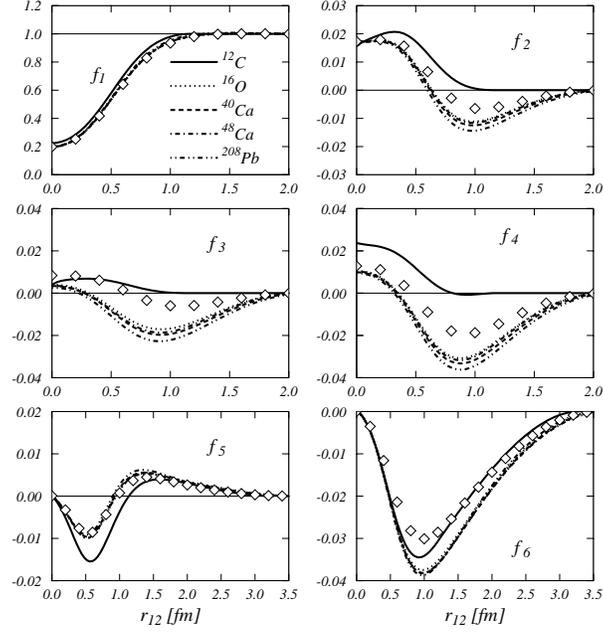}
\caption{\small Two-body correlation functions $f_p$, as a function of
 the internucleonic distance. The symbols indicate the correlation
 functions obtained for symmetric nuclear matter (see text for
 explanations). 
}
\label{fig:corr} 
\end{figure}

The two-body correlation functions obtained for the $v'_8$ plus $UIX$
interaction are shown in Fig. \ref{fig:corr} for each channel. In the
figure, we also show with the diamonds the nuclear matter correlations.
These last correlations are obtained with the Euler procedure by using
the same healing distances used for finite nuclei but with constant
density value of the nuclear matter the saturation point.  Analogous
results have been obtained for the $v_{14}$ plus $UVII$ interaction.

We would like to point out two features of these results. The first
one, is that the various correlations are rather similar for all the
nuclei considered.  Only the \car results are out of this general
trend. This fact indicates that the correlation functions are more
sensitive to the characteristics of the nucleon-nucleon interaction
than to the shell effects.  The second point we want to mention is the
difference between the {\sl healing distances} of the central and
tensor channels.  In the two-body interaction, the tensor channels are
active at larger distances with respect to the central channels, and
this produces the consequences on the correlation functions we have
just pointed out.  It is interesting to notice that the {\sl healing}
for the tensor channels appears at distances larger than the
experimental charge radius of the deuteron.

In this discussion the \car nucleus remains an anomaly. The $f_1$
correlation is rather similar to those of the other nuclei, all the
other two-body correlations are remarkably different, especially $f_3$
and $f_4$ related to the spin and spin-orbit dependent terms. 

Also the nuclear matter results are not on the band formed by the
correlations found for the \oxi - \lead nuclei. 
The differences are
not so large as those of the \car, but still rather evident. It is
remarkable the similarity in the $f_1$ channel and in the $f_5$
channel, while all the other channels show noticeable differences.

Our calculations are organized in such a way that we first solve the
FHNC equation for the scalar correlation only, and in a second step we
solve the FHNC/SOC equations with the other state dependent
correlation terms. The validity of our solution techniques, from both
the theoretical and numerical points of view, can be appreciated by
observing the convergence in the density sum rules shown in Tab.
\ref{tab:sumr}.  In this table we give the normalization of the one-
and two-body densities (\ref{eq:obdc}) and (\ref{eq:tbdc}). The $S_1$
and $S_2$ sum rules have been defined in such a way that all of them
are normalized to unity. The results of the table show that the
Jastrow-FHNC calculations, those with $f_1$ only, produce densities
deviating by few parts on a thousand from the correct normalization.
The inclusion of state dependent terms in the correlations, the
FHNC/SOC calculations, is responsible for an increase of these
deviations up to a maximum value of 5.5 \%. In many-body systems
saturated in spin and isospin one can also test spin and isospin
density sum rules. In the systems we are investigating these sum rules
are not any more valid.

The main result of our paper is summarized in Tab. \ref{tab:bene}
where we give the values of the binding energies per nucleon for all
the five nuclei considered, and we compare them with the experimental
binding energies \cite{aud93}. In the table we present the various
terms contributing to the total energy: the kinetic energy $T$, the
two-body interaction, where the the contribution of the first six
channels $V^{6}_{2-body}$ and that of the spin-orbit interaction
$V_{LS}$ are separately given, the Coulomb interaction $V_{Coul}$ and
the three-body force $V_{3-body}$. In the kinetic energy term the
contribution of the center of mass motion, calculated as discussed in
\cite{co92}, has already been subtracted. 

The various terms show some saturation properties.  For example the
values of kinetic energies per nucleon $T$ increase up to \caI and
then they remain almost stable around a value of about 40 MeV.  An
analogous behaviour is shown by the $V^{6}_{2-body}$ terms.  The
contribution per nucleon increases with increasing number of nucleons
up to \caI. After that it is evident that saturation appears in
heavier systems.

We have already mentioned the fact that the spin-orbit terms are not
treated consistently in the FHNC/SOC computational scheme, but they
are evaluated by using some approximation. In any case, in all the
nuclei considered, their contributions are of the order of few percent
with respect to the $V^{6}_{2-body}$ contributions. As a further test,
we have done calculations in \oxi and \caI switching off the
spin-orbit terms in the mean field potential. In this case the
spin-orbit partner single particle wave functions are identical. The
differences in the total spin-orbit contributions, with respect to the
values given in Tab. \ref{tab:bene} are within the numerical
uncertainty.

The contribution of the Coulomb term $V_{Coul}$ is evaluated within
the complete FHNC/SOC computational scheme.  As expected, the
behaviour with increasing size of the nucleus does not show saturation
because of the long range nature of the interaction. The Coulomb terms
behave as expected, their contributions increase with increasing
number of protons.  The apparent inversion of this trend from \caI to
\caII is due to the representation in terms of energy per nucleon,
which in this case is misleading, since the proton number is the same
for the two nuclei. In this case it is better to compare the total
values of the Coulomb energies, 78.40 MeV for \caI and 75.36 for
\caII. The 3.8\% difference between these to values is due to the
different structure of the two nuclei. The inclusion of the Coulomb
repulsion reduces the nuclear binding energies.  To simplify the
reading of the table we give in the raw labeled $T+V(2)$ the
intermediate energy value obtained with the two-body force only,
Coulomb interaction included.  The total binding energies without the
three-body forces are all above the experimental values.

In addition to that, there is the contribution of the three-body
force.  As discussed in \cite{fab00}, the two terms composing this
interactions provide contributions of different sign, the
Fujita-Miyazawa term is attractive, while the other term is repulsive.
In our calculations the total contribution of the three-body forces is
always globally repulsive.

The comparison with the experimental energies indicates a general
underbinding of about 3-4 MeV per nucleon.  In the general behaviours
we have discussed so far, the anomaly of the \car nucleus emerges. In
all our calculations the attractive contribution of the $V^6_{2-body}$
is relatively smaller than for the other nuclei. As a consequence,
this nucleus results barely bound in our calculations.  Some crucial
physics ingredient relevant in \car while negligible for the other
other nuclei is missing in our approach. Probably this has to do with
soft deformations of the \car nucleus which we are unable to treat.

The comparison between the two interactions indicates that the
$v_{14}$ plus $UVII$ is more attractive than the $v'_8$ plus $UIX$.
This fact is already present when the two-body interaction only is
considered. The situation is enhanced by the three-body force, more
repulsive for the $UIX$ case than for the $UVII$ one. The
contributions of the spin-orbit term in the two cases have different
sign, they are attractive for $v'_8$ and slightly repulsive for
$v_{14}$. The differences in the total energies calculated with the
two sets of interactions go from a minimum of 5\% (\oxi) to a maximum
of 18\% (\lead).

Despite the remarkable differences in the energy, the two interactions
produce very similar results in the observables related to the
density distributions. For this reason, henceforth, we shall present
results obtained with the $v'_8$ plus $UIX$ interaction only.

\begin{figure}[ht]
\includegraphics [scale=0.45]
                 {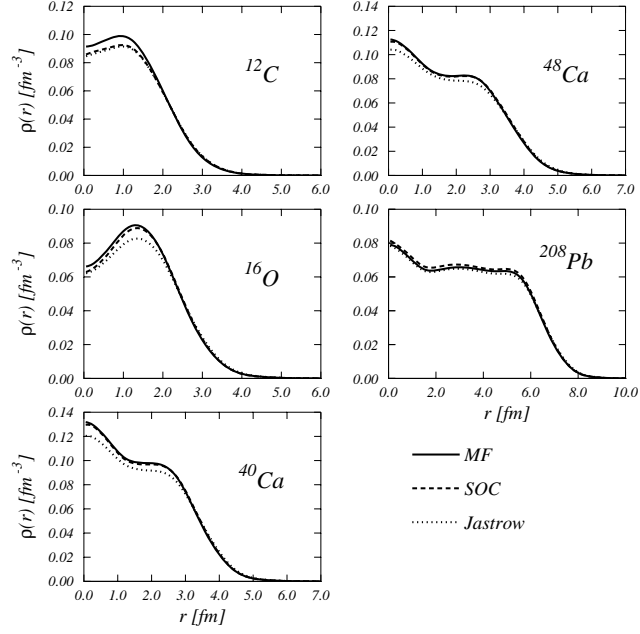}
\caption{\small Proton density distributions. The full lines are
the mean field distributions, the dotted ones have been obtained by
using scalar correlations ($f_1$) only, and the dashed lines show the
results of the complete calculation.
}
\label{fig:densp} 
\end{figure}

\begin{figure}[ht]
\includegraphics [scale=0.45]
                 {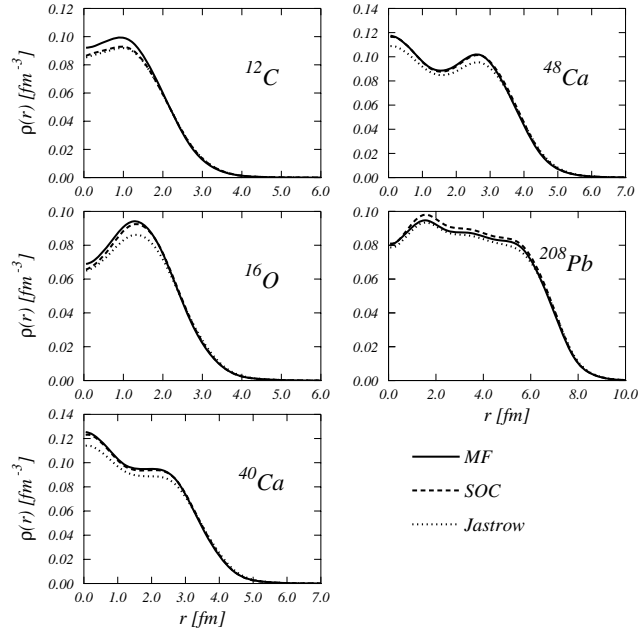}
\caption{\small The same as Fig. \protect\ref{fig:densp}, for the
                 neutron density distributions.
}
\label{fig:densn} 
\end{figure}

In Figs. \ref{fig:densp} and \ref{fig:densn} we show the one-body
density distributions for protons and neutrons respectively.  The full
lines show the uncorrelated distributions, the dotted lines those
obtained by using scalar correlations only $f_1$, we call Jastrow
these calculations, and the dashed lines the results of the complete
calculation.  The Jastrow results produce distributions which are
smaller at the center of the nucleus with respect to the mean-field
distributions.  This effect is strongly reduced when all the
correlations are included in the calculation.  These findings are in
agreement with the results of ref. \cite{ari97} where a first-order
cluster expansion has been used. It is evident the scarce relevance of
short-range correlations in reducing the occupation probability of the
3s1/2 proton state in \lead.  As pointed out in \cite{ang01}, the
effect of long-range correlations seems to be main reduction
mechanism.

\begin{figure}[ht]
\includegraphics [scale=0.45]
                 {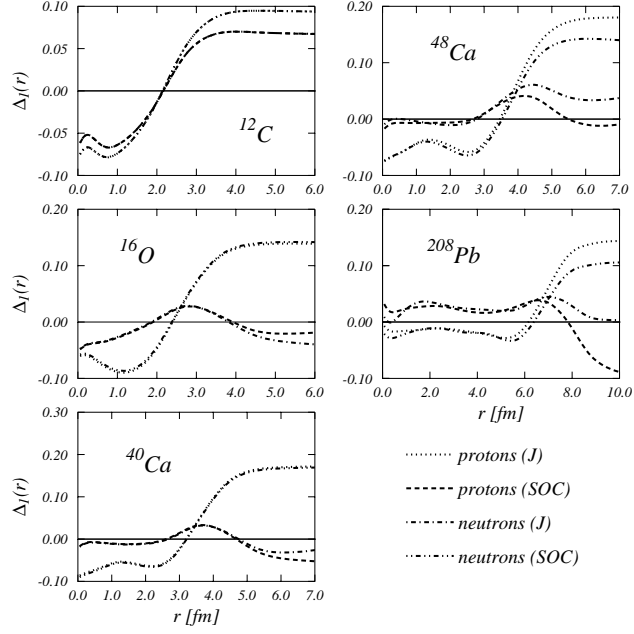}
\caption{\small 
  Normalized ratios (\ref{eq:delta1}) for proton and neutron
  distributions. The meaning of the lines is given in the legend,
  where we have indicated with $J$ the results obtained with scalar
  correlation ($f_1$) only, and with $SOC$ the results of the full
  calculation.  
}
\label{fig:densd} 
\end{figure}

In order to investigate the relative effects of the various
correlations on the protons and neutrons distributions we studied the
ratio:
\begin{equation}
\Delta_1(r) = \frac{\rho(r) - \rho^o(r)}{\rho^o(r)} 
\label{eq:delta1}
\end{equation}
where, as it has been defined in eq. (\ref{eq:obdu}), $\rho^o(r)$ is
the uncorrelated, mean-field, density. The $\Delta_1$ ratios are shown
in Fig. \ref{fig:densd} for all the nuclei under investigation.
The behaviour of the various lines indicates
more similarity between results obtained with the same type of
correlation than between protons and neutrons. This result is quite
remarkable because of the large difference between proton and neutron
distributions, especially for \caII and \lead.

\begin{figure}[ht]
\includegraphics [scale=0.45]
                 {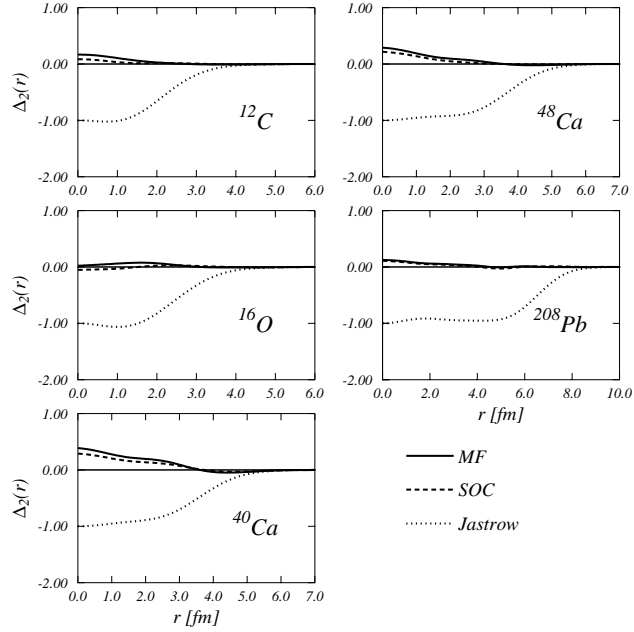}
\caption{\small Normalized ratios (\ref{eq:delta2}) for charge density
distributions. The meaning of the lines is given in the legend.
}
\label{fig:chd} 
\end{figure}

A direct comparison with the empirical charge density distributions is
not very meaningful since the mean-field potential has already been
fixed to reproduce, at its best, these distributions. The only
statements we can make is whether the inclusion of the correlations
improves or not the agreement with the data. For this reason we show
in Fig.  \ref{fig:chd} the ratios
\begin{equation}
\Delta_2(r) = \frac{\rho(r) - \rho_{exp}(r)}{\rho_{exp}(0)} 
\label{eq:delta2}
\end{equation}
where the experimental charge distributions have been taken from
\cite{dej87}.
The inclusion of correlations lowers the charge density distributions at
the center of the nuclei. If scalar correlation only are considered
the effect is too large. When all the correlations are included the
effect is strongly reduced and the agreement with the empirical
densities improves.

The effects of the short-range correlations are better shown on the
two-body density matrices (\ref{eq:tbdc}). We define the operator
dependent two-body density matrices as:
\begin{eqnarray}
\nonumber
&~& \rho^{\alpha \beta,p}_2(\br_1,\br_2) = 
\frac {A(A-1)} {<\Psi|\Psi>} \\
&~& \int dx_3...dx_A \,\Psi^*(x_1,x_2,x_3,...,x_A)
{\cal P}^{\alpha}(1) O^p(x_1,x_2) {\cal P}^{\beta}(2)
\Psi(x_1,x_2,x_3,...,x_A)
\label{eq:otbdm}
\end{eqnarray}
where we have indicated with ${\cal P}^\alpha(i)$ the operator selecting
protons or neutrons and we have understood the trace on the 
spins and isospins  of the particles 1 and 2. 
We calculate the values of (\ref{eq:otbdm}) in
terms of relative distance $r_{12}=|\br_1 - \br_2|$
between the two considered nucleons
\begin{equation}
\rho^{\alpha \beta,p}_2(r_{12}) = 
\int d {\bf R}_{12} \rho^{\alpha \beta,p}_2(\br_1,\br_2)
\label{eq:rtbdm}
\end{equation}
where ${\bf R_{12}}=(\br_1 + \br_2)/2$ is the center of mass coordinate of
the pair. In the FHNC/SOC computational scheme, we calculated the
two-body density matrices as:
\begin{eqnarray}
&~& \rho^{\alpha\beta,2k_3-1+l_3}_2(\br_1,\br_2) =  
\frac{f_{2k_1-1+l_1}(r_{12})f_{2k_2-1+l_2}(r_{12})}
{f_{1}^2(r_{12})}\Bigg\{
I^{k_1k_3k_2}A^{k_2}\chi^{\alpha \beta}_{l_1+l_2+l_3}
\rho^{\alpha \beta}_{2,dir}(\br_1,\br_2)\\
\nonumber
&~&+\Big(I^{k_4k_1k_5}I^{k_2k_3k_5} 
A^{k_5} \rho_{2,exc}^{\alpha \beta}(\br_1,\br_2)
+I^{k_4k_1k_5}I^{k_2k_3k_6} I^{k_5 k_6 2} 
\rho_{2,excj}^{\alpha \beta}(\br_1,\br_2)  \Big)
\Delta^{k_4} \sum_{l_4=0}^1\chi^{\alpha \beta}_{l_1+l_2+l_3+l_4}\Bigg\}
\label{eq:soctbdm}
\end{eqnarray}
where a sum on the repeated indexes is understood and the symbols of
eq. (\ref{eq:w02}) have been used.

\begin{figure}[ht]
\includegraphics [scale=0.45]
                 {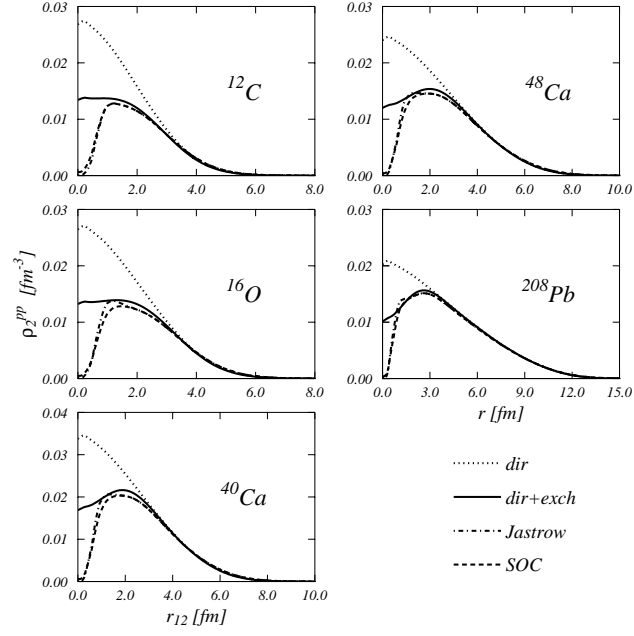}
\caption{\small 
Proton-proton scalar (p=1) two-body distributions 
(\protect\ref{eq:rtbdm})
as a function of the internucleonic distance. The dotted lines show
the uncorrelated two-body distributions calculated by considering
only the first term of eq. (\protect\ref{eq:tbdu}), the direct
one. The full lines show the total uncorrelated two-body distributions
as given by eq. (\protect\ref{eq:tbdu}). The other two types of lines
indicate the results obtained with scalar correlations only (Jastrow) 
and in the complete calculation (SOC).
}
\label{fig:tbdpp} 
\end{figure}

\begin{figure}[ht]
\includegraphics [scale=0.45]
                 {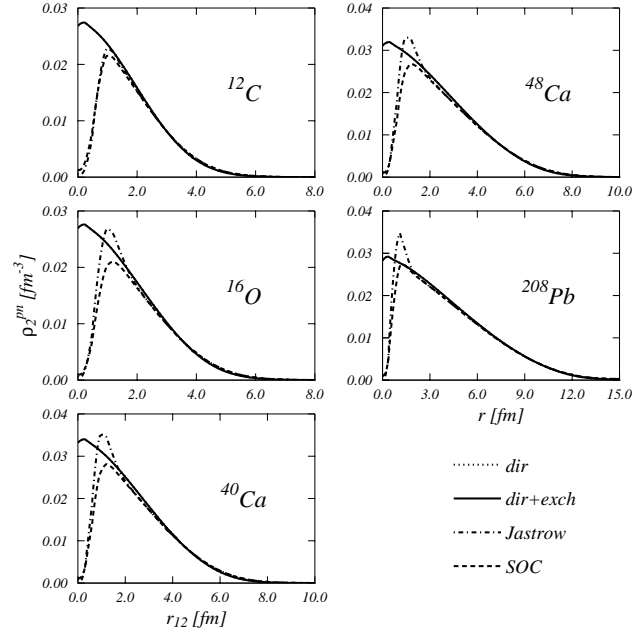}
\caption{\small 
The same as Fig. \protect\ref{fig:tbdpp} for the proton-neutron
case. The dotted and full lines coincide, since the exchange term of
eq. (\protect\ref{eq:tbdu}) is zero.
}
\label{fig:tbdpn} 
\end{figure}

\begin{figure}[ht]
\includegraphics [scale=0.45]
                 {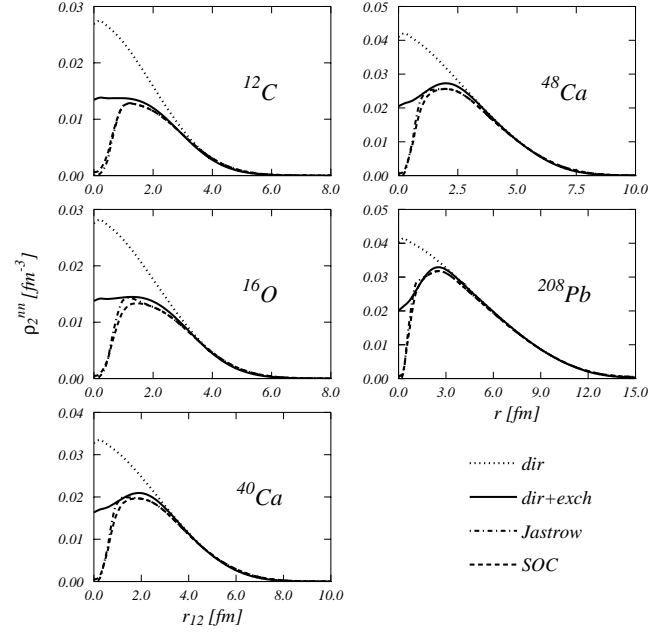}
\caption{\small 
The same as Fig. \protect\ref{fig:tbdpp} for the neutron-neutron.
}
\label{fig:tbdnn} 
\end{figure}

Let's consider first the Jastrow case, where only $f_1$ correlation
functions are present, i.e. $2k_1-1+l_1=2k_2-1+l_2=1$. In this case,
apart from a normalization factor, eq. (\ref{eq:otbdm}) gives the
probability density of finding two nucleons with isospin
third-components $\alpha$ and $\beta$ at the distance $r_{12}$.  These
scalar two-body density functions, or probability densities, are shown
in Figs. \ref{fig:tbdpp}, \ref{fig:tbdpn} and \ref{fig:tbdnn}, for all
the nuclei we have investigated.  The dotted lines represent the
uncorrelated joint probability densities of finding the two nucleons
as often given in the literature, see for example \cite{ben93}.  This
joint probabilities are obtained as product of the uncorrelated
one-body densities.  This definition of the uncorrelated two-body
density can be meaningful from the probabilistic point of view, but it
is misleading in our framework, since it corresponds to use only the
first, direct, term of eq. (\ref{eq:tbdu}). In fermionic systems, the
uncorrelated two-body density is given by the full expression
(\ref{eq:tbdu}) containing also the exchange term. These complete
uncorrelated two-body densities, are shown in Figs.  \ref{fig:tbdpp},
\ref{fig:tbdpn} and \ref{fig:tbdnn} by the full lines. The effects of
the short-range correlations can be deduced by comparing these lines
with the dashed-dotted lines obtained with the $f_1$ scalar
correlations only, we call these Jastrow results, and with the dashed
lines showing the full FHNC/SOC results.

In all our results, the correlations reduce the values of the two-body
densities at short internucleon distances. The exchange term of the
uncorrelated density, already contributes to this reduction, but the
major effect is produced by the short-range correlations, and mainly
by the scalar correlations.  The effects of nucleon-like two-body
densities are rather similar, as it can be seen comparing Figs.
\ref{fig:tbdpp} and \ref{fig:tbdnn}. When the pair of nucleons is
composed by different particles the situation slightly changes. Beside
to the strong reductions at small distances, the correlations produce
enhancements around 2 fm in all the nuclei considered.

\begin{figure}[ht]
\includegraphics [scale=0.45]
                 {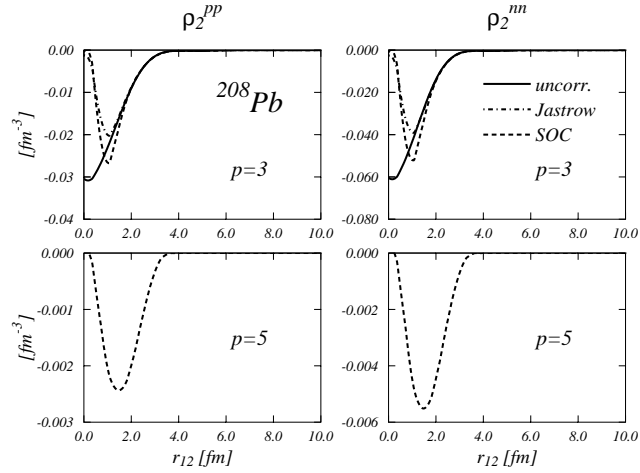}
\caption{\small 
  Proton-proton and neutron-neutron two-body distribution functions
  (\protect\ref{eq:rtbdm}) of \lead nucleus. As discussed in the text,
  for equal particles, the p=2,4,6 distribution functions are
  respectively equal to the p=1,3,5 ones.
}
\label{fig:leadpp} 
\end{figure}

\begin{figure}[ht]
\includegraphics [scale=0.45]
                 {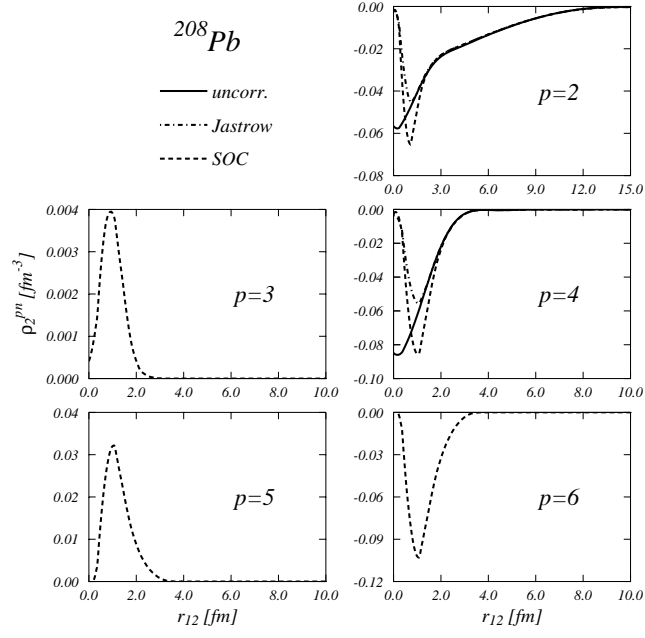}
\caption{\small 
The same as Fig. \protect\ref{fig:leadpp} for the proton-neutron case.
}
\label{fig:leadpn} 
\end{figure}

In order to discuss the effects of the correlations on the other
operator dependent two-body densities (\ref{eq:otbdm}), we show in
Figs. \ref{fig:leadpp} and \ref{fig:leadpn} the results for the \lead
nucleus only. The results of the other nuclei considered have similar
features \cite{bis05}. In these figures, we used the same line
conventions of Figs. \ref{fig:tbdpp}, \ref{fig:tbdpn} and
\ref{fig:tbdnn}.  When the external particles are nucleons of the same
type, i.e.  $\alpha=\beta$ in eqs. (\ref{eq:otbdm}) -
(\ref{eq:soctbdm}), the isospin trace in eq. \ref{eq:otbdm} is one.
In this case, the p=1,2,3 two-body densities are equal, respectively,
to those calculated with p=2,4,6.  The same idea can be expressed more
formally by stating that, in eq.  (\ref{eq:soctbdm}) for
$\alpha=\beta$ we have $\rho^{\alpha \alpha,2k-1}_2(r_{12}) =
\rho^{\alpha \alpha,2k}_2(r_{12})$, since $\chi_n^{\alpha\alpha}=1$
for every value of $n$.  For this reason, in Fig. \ref{fig:tbdpp}, we
present only the p=3 and p=5 two-body densities for both proton-proton
and neutron-neutron cases. In the proton-neutron case the two-body
densities are different in every channel.

We first notice that tensor two-body densities, p=5, and also p=6 in
Fig. \ref{fig:tbdpn}, are different from zero only in the full
FHNC/SOC calculations.  This occours because in order to get a spin
trace different from zero in eq. \ref{eq:otbdm} at least two tensor
operators are needed.  In Fig. \ref{fig:tbdpn} also the p=3 spin
density is different from zero only in the FHNC/SOC case.  The reason
is that, in the calculation of this density,for the uncorrelated and
the Jastrow cases, the first term of eq. (\ref{eq:tbdu}), the direct
one, is zero because of the spin trace, while the exchange term is
zero because of the different isospin third component.

We observe from Figs. \ref{fig:tbdpp} and \ref{fig:tbdpn} that the
effects of the correlations are analogous to those already discussed
for the scalar two-body densities: the values of the densities is
strongly reduced at short internucleonic distances. In addition to
that it is worth to mention the change of sign in the isospin, p=2,
densities in the proton-neutron case.  The most remarkable feature is
the range of the various densities. The scalar densities of Figs.
\ref{fig:tbdpp}, \ref{fig:tbdpn} and \ref{fig:tbdnn} and the isospin
densities, p=2, extend up to relative distances comparable to the
dimensions of the nucleus, 15 fm. On the contrary, all the other
two-body densities have much smaller ranges, of the order of 3-4 fm.
These features are present in all the nuclei considered \cite{bis05}.

\section{SUMMARY AND CONCLUSIONS}
\label{sec:summary}
With this paper we made an important step forward in our project aimed
to provide a microscopical description of medium-heavy nuclei, by
using the CBF theory.  With respect to the works of Refs.
\cite{fab98,fab00} we have applied for the first time the FHNC/SOC
computational scheme to nuclei with different number of protons and
neutrons and in the $jj$ coupling scheme.  We have used two different
sets of microscopic interactions: the Argonne $v'_8$ implemented with
the Urbana IX three-body force, and the truncated Urbana $v_{14}$ plus
the Urbana VII three-body force.  Binding energies have been presented
in Tab. \ref{tab:bene} and this is the main result of our work.  With
the noticeable exception of the \car nucleus, our binding energies are
about 3-4 MeV/nucleon above the experimental ones, consistently with
our previous results \cite{fab00} and with those obtained in nuclear
matter.

In our calculations we did not include the contributions of the
spin-orbit terms of the correlation and of the p$>$ 8 terms of the
two-body potential, which in \cite{fab00} have been estimated by means
of a local density interpolation of the nuclear matter results. These
contributions are attractive and of the order of 1 MeV/nucleon. This
produces differences with the experimental values of the order of
those presented in \cite{fab00} where only the \oxi and \caI nuclei
were considered.  A number of possible sources of the remaining
disagreement should be investigated such as the three-body
correlations and perturbative corrections of the two-body
correlations.

The \car nucleus, strongly underbound, stands out of this general
trend.  As we have already said some specific characteristics of this
nucleus, not relevant in the other nuclei we have considered, is not
properly treated in our calculations. We think this could be the
presence of deformation phenomena.

Short-range correlations produce small effects on the one-body density
distributions.  The distributions at the center of the nucleus are
lowered by the correlations. This effect is relatively large when only
scalar correlations are used, and it is strongly reduced when the
other terms of the correlations are included.  These results are in
agreement with the findings of \cite{ari97} where a simple first-order
expansion of the density has been used.

The study of the two-body densities show that the short-range
correlations strongly reduce the probability of finding two nucleons
at distances smaller than 0.5-0.6 fm. The scalar and isospin,
two-body densities, extend on the full nuclear volume, while the other
densities have a maximum range of about 3 fm.

Our work shows that microscopic variational calculations for
medium-heavy nuclei have reached the same degree of accuracy as
nuclear matter calculations. Our approach provides a good description
of the short-range correlations, but it shows some deficiency related
to the proper inclusion of long-range correlations. The next obvious
step would be to extend the theory to have a better treatment of these
last type of correlations.

%
%

%
\clearpage
\newpage 
%
%
\begin{table}[h]
\begin{center}
\begin{tabular}{l|rrrrrr}
\hline
  &       & \car & \oxi & \caI & \caII & \lead \\
\hline
$v'_8$ +   & $d_c$   & 1.20 & 2.10 & 2.15 & 2.10 & 2.20 \\
$UIX$      & $d_t$   & 3.30 & 3.70 & 3.66 & 3.70 & 3.60 \\
\hline
$v_{14}$ + & $d_c$   & 1.40 & 2.10 & 2.15 & 2.10 & 2.20 \\
$UVII$     & $d_t$   & 3.30 & 3.80 & 3.86 & 3.90 & 3.80 \\
\hline
\end{tabular}
\end{center} 
\caption{\label{tab:heal} 
\small Values, in fm, of the {\sl healing distances}, obtained
minimizing the energy functional (\protect\ref{eq:efun}) with the 
interactions we have used in our calculations. 
}
\end{table}
%
%
\begin{table}[h]
\begin{center}
\begin{tabular}{llccccc}
\hline
       &  & \car & \oxi & \caI & \caII & \lead \\
\hline
 $S^p_{1}$ &$(J)$     & 1.000 & 1.000 & 1.000 & 1.000 & 0.999 \\
 $S^p_{1}$&$(SOC)$   & 0.997 & 1.006 & 1.008 & 1.013 & 1.024 \\
 $S^n_{1}$&$(J)$     & 1.000 & 1.000 & 1.000 & 0.999 & 0.999 \\
 $S^{n}_{1}$&$(SOC)$ & 0.997 & 1.006 & 1.008 & 1.021 & 1.027 \\
 $S_{2}$&$(J)$       & 1.004 & 1.003 & 1.001 & 1.000 & 0.998 \\
 $S_{2}$&$(SOC)$     & 0.996 & 1.025 & 1.022 & 1.042 & 1.055 \\
\hline
\end{tabular}
\end{center} 
\caption{\label{tab:sumr} 
\small Normalization of the one- and two-body densities $S_1$ and
$S_2$ respectively. For the one-body densities we give separately the
contributions of the protons ($p$) and neutrons ($n$).  
The label $J$ (Jastrow) indicates the calculation done with the scalar,
$f_1$, correlation only, while the $SOC$ labels indicate the results
of the complete calculation. 
}
\end{table}
%
%
%
\begin{table}[ht]
\begin{center}
\begin{tabular}{l|lrrrrrr}
\hline
   &              & \car & \oxi  & \caI & \caII & \lead \\
\hline
& $T$              &  27.13 &  32.33 &  41.06 &  39.64 &  39.56  \\ 
& $V^{6}_{2-body}$ & -29.13 & -38.15 & -48.97 & -46.60 & -48.43  \\ 
& $V_{LS}$         &  -0.25 &  -0.38 &  -0.39 &  -0.35 &  -0.45  \\ 
$v'_8$ + & $V_{Coul}$       &   0.67 &   0.86 &   1.96 &   1.57 &   3.97  \\
$UIX$    & $T+V(2)$         &  -1.58 &  -5.34 &  -6.34 &  -5.74 &  -5.35  \\
& $V_{3-body}$     &   0.66 &   0.86 &   1.76 &   1.61 &   1.91  \\ 
& $E$              &  -0.91 &  -4.49 &  -4.58 &  -4.14 &  -3.43  \\ 
\hline
& $T$               & 24.63 &  29.25 & 37.70  & 36.47  &  36.48 \\
& $V^{6}_{2-body}$  &-27.08 & -35.84 &-47.16  &-44.86  & -46.87 \\
& $V_{LS}$          &  0.05 &   0.03 &  0.07  &  0.09  &   0.04 \\ 
$v_{14}$ + & $V_{Coul}$        &  0.68 &   0.88 &  2.02  &  1.59  &   4.03 \\
$UVII$     & $T+V(2)$          & -1.72 &  -5.68 & -7.37  & -6.71 &   -6.32  \\
& $V_{3-body}$      &  0.54 &   0.69 &  1.28  &  1.15  &   1.41 \\  
& $E (UIX)$         & -1.18 &  -4.99 & -6.09  & -5.56  &  -4.91 \\ 
\hline
&  $E_{exp}$        & -7.68   &  -7.97 & -8.55   &  -8.66 &  -7.86  \\ 
\hline
\end{tabular}
\caption{\label{tab:bene}
\small Energies per nucleon in MeV. We have indicated with $T$ the
kinetic energy, with $V^{6}_{2-body}$ the contribution of the first
six channels of the two-body interaction, with $V_{LS}$ the spin-orbit
contribution, with $V_{Coul}$ the contribution of the Coulomb
interaction and with $V_{3-body}$ the total contribution of the
three-body force. The raws labeled $T+V(2)$ show the energies
obtained by considering the two-body interactions only.
The experimental values are taken form
\protect\cite{aud93}. 
}
\end{center} 
\end{table}
%
%
\clearpage
\newpage
%
%
%
%
%
%
%
%
%
%
%
%
%
%
%
%
%
%
%
%
%
%
\end{document}